# Coulomb blockade in Etched Single and Few Layer MoS$_2$ Nanoribbons

*Dharmraj Kotekar-Patil[1*], Jie Deng[1], Swee Liang Wong[1,2], Kuan Eng Johnson Goh[1,2§]*

1. Institute of Materials Research and Engineering, Agency for Science Technology and Research, 2 Fusionopolis Way, 08-03 Innovis, 138634, Singapore

2. Department of Physics, National University of Singapore, 2 Science Drive 3, 117551, Singapore

**ABSTRACT**

Confinement in two-dimensional transition metal dichalcogenides is an attractive platform for trapping single charge and spins for quantum information processing. Here, we present low temperature electron transport through etched 50-70nm MoS$_2$ nanoribbons showing current oscillations as a function of gate voltage. On further investigations current through the device forms diamond shaped domains as a function of source-drain and gate voltage. We associate these current oscillations and diamond shaped current domains with Coulomb blockade due to single electron tunneling through a quantum dot formed in the MoS$_2$ nanoribbon. From the size of the Coulomb diamond, we estimate the quantum dot size as small as 10-35nm. We discuss the possible origins of quantum dot in our nanoribbon device and prospects to control or engineer the quantum dot in such etched MoS$_2$ nanoribbons which can be a promising platform for spin-valley qubits in two-dimensional transition metal dichalcogenides.

**INTRODUCTION**

Transition metal dichalcogenides (TMDCs) monolayers are atomically thin materials which have attracted enormous interest due to its remarkable optical and electrical properties. Along with its direct bandgap, it also possesses strong spin-orbit coupling and spin-valley coupling which makes it an attractive research platform for spin based physics. Due to its naturally ultrathin layers, TMDCs form an ideal material system for confining single charge and spins for spin based quantum information processing.

Molybdenum disulfide (MoS$_2$) is one of the most well studied material among the TMDC class. A large body of work shows that decently low resistance ohmic contacts with low Schottky barrier are achievable [1-7]. A first step towards spin based quantum information processing would be demonstration of charge and spin confinement. Signatures of conductance quantization [8,9] and formation of single quantum dot and double quantum dot has been demonstrated in single layer MoS$_2$ [10-12] and other TMDC materials [13-15]. These confined structures have been created using split-gate technique with top electrostatic gates using either hexagonal-boron nitride (hBN) encapsulated TMDC [8-11,13-15] or atomic layer deposition (ALD) grown gate dielectric [16]. Typical quantum dot size achieved with such a technique is on the order of few hundreds of nanometers and require several gates to confine single quantum dot [10,15]. A recent report claims that the effective mass measured in a single layer MoS$_2$ [10] is higher than the theoretically predicted [16] value, implying the necessity of a stronger confinement (10-50nms) to reach few electron regime. To achieve such a strong confinement in commonly used split-gate technique requires several gates with a very small gate pitch which is quite challenging.

In this letter, we present the first low temperature transport spectroscopy measurements on single and few layer $MoS_2$ nanoribbons (NR). We utilize an alternative approach where instead of creating a confinement by multiple top-electrostatic gates, we create a confinement by etching the $MoS_2$ flake into a NR of width (W) = 50-70nm and length (L) = 500nm. In such a NR device, we show that at low temperature (3K) a quantum dot is formed, exhibiting clear Coulomb blockade via Coulomb oscillations and diamonds. From the shape of the Coulomb diamond, we estimate the size of the quantum dot to be ~10-35nm. Size of the quantum dot is comparable to the dimensions of the NR. Finally, we discuss the origin of the quantum dot in our NR device and ways to control and engineer the quantum dot for future work towards spin based quantum information processing.

**FABRICATION**

Device fabrication started by transferring mechanically exfoliated $MoS_2$ flake on to a heavily doped silicon substrate with a 300nm $SiO_2$ layer (figure 1a and 1d). Thickness (in terms of number of layers) of the transferred flakes were identified first by optical contrast and then by photoluminescence and Raman spectroscopy (described later) to further verify the flake thickness especially for monolayer $MoS_2$. The silicon substrate with $MoS_2$ flakes was then spin coated with PMMA and electron beam lithography was used to define the source and the drain contacts. Ti/Au (= 10/80 nm thick) was then evaporated to form ohmic contact with $MoS_2$. Sufficiently large Ti/Au metal contacts (4-5 microns) overlapping the $MoS_2$ was defined, much larger than the transfer length of electrons [17] into $MoS_2$ layer to ensure efficient charge injection. To remove the residual resist from the $MoS_2$ surface and further improve the contact resistance, the device was annealed in H$_2$/Ar$_2$ (10/90%) atmosphere at 200°C for 2 hours. To etch out a nanoribbon (NR), PMMA was spin coated on the MoS$_2$ and an electron beam lithography was used to define a NR pattern such that only the desired area of NR region was protected by PMMA during SF$_6$ dry plasma exposure. After performing the etching process, the chip was immersed in acetone to remove the PMMA which leaves a MoS$_2$ NR between the source and drain metallic contacts (figure 1c and 1f).

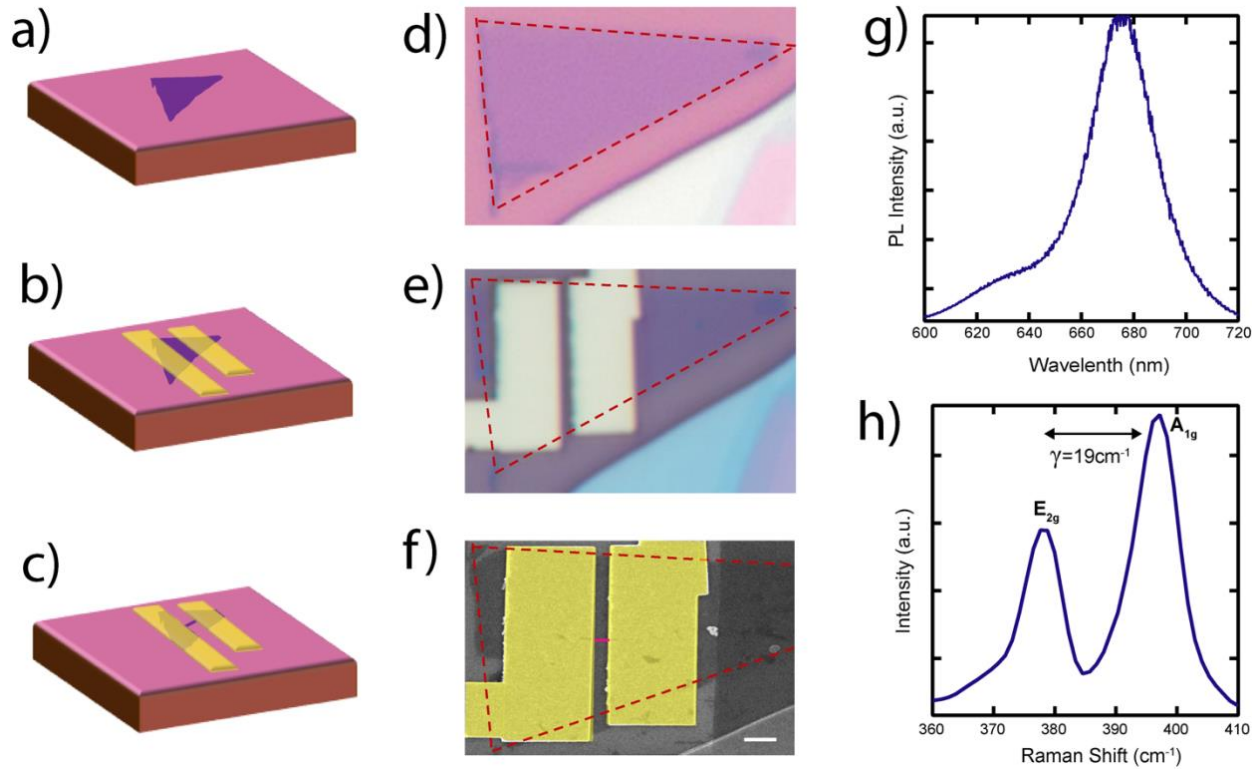

*Figure 1. Schematic of fabrication process for NR based field effect transistor in single layer MoS$_2$ (a-c). Schematic of the device fabrication showing step-by-step process going from a MoS$_2$ flake (triangle) to nanoribbon (NR) device where yellow stripes are the metallic contacts. (d-f) actual device images showing step-by-step process corresponding (a-c). Scale bar in panel (f) is 1μm. (g, h) shows PL and Raman spectroscopy respectively to validate single layer MoS$_2$.*

Optical characterizations were performed before fabricating the device to identify monolayer MoS$_2$ by photoluminescence (PL) and Raman spectroscopy. Figure 1g and 1h show the room temperature photoluminescence measurement and Raman spectroscopy respectively. A 532nm laser was used to perform PL and Raman spectroscopy. Figure 1g shows a high intensity peak centred at 670 nm suggesting the presence of a direct bandgap, a typical feature of single layer

MoS$_2$. This value of 670nm corresponds to a bandgap of 1.8eV, in agreement with the bandgap of single layer MoS$_2$ [18]. Single layer MoS$_2$ was further verified by Raman spectroscopy exhibiting two characteristic A$_{1g}$ and E$_{2g}$ mode separated by $\Delta\gamma$=19 cm$^{-1}$ [19,20]. Multilayer and single layer MoS$_2$ flakes were distinguished from the PL measurements where multilayer flakes exhibited multiple emission peaks with significantly weaker PL intensities.

Electrical measurements presented in this letter were performed by measuring the two-point source-drain current (I$_d$) as a function of source – drain voltage (V) and global backgate voltage (V$_{bg}$). Room temperature and 77K electrical characterisation of the devices were performed on a Janis probe station under vacuum (10$^{-4}$ mTorr) whereas low temperature measurements were performed in a Janis cryostat at 3K.

**RESULTS**

In this letter, we present data measured on two devices: one on a single layer MoS$_2$ (D1, figure 2) and another on few layer MoS$_2$ (D2, figure 3) NR transistor.

Figure 2a shows three traces of source-drain current measured through device D1 at V=100mV as a function of V$_{bg}$. The red trace (•) corresponds to the transconductance measurement at a temperature (T) =300K before etching the single layer MoS$_2$ flake (W=6µm and L=500nm) into a NR. The pinch-off voltage is significantly negative (V$_{bg}$ ~ -80V) consistent with other reports [21]. At V$_{bg}$= 60V, I$_d$ reaches 4µA which corresponds to a total resistance of 25KΩ (channel resistance + 2 x contact resistance). From the linear part of I$_d$-V$_{bg}$, the field effect mobility can be estimated using the $\mu = \frac{L}{WC_{ox}V}\frac{dI_d}{dV_{bg}}$, where L is the channel length, W is the channel width and C$_{ox}$ is the gate capacitance. The mobility extracted from the red trace in figure 2a is about 2cm$^2$/V.s. The purple trace (Δ) in figure 2a shows transconductance measurement at T=300K after etching the single layer MoS$_2$ flake into a nanoribbon of dimensions: W=50nm and L=500nm. Threshold voltage for etched NR device shifted significantly to the right, consistent with the previous reports [22,23].

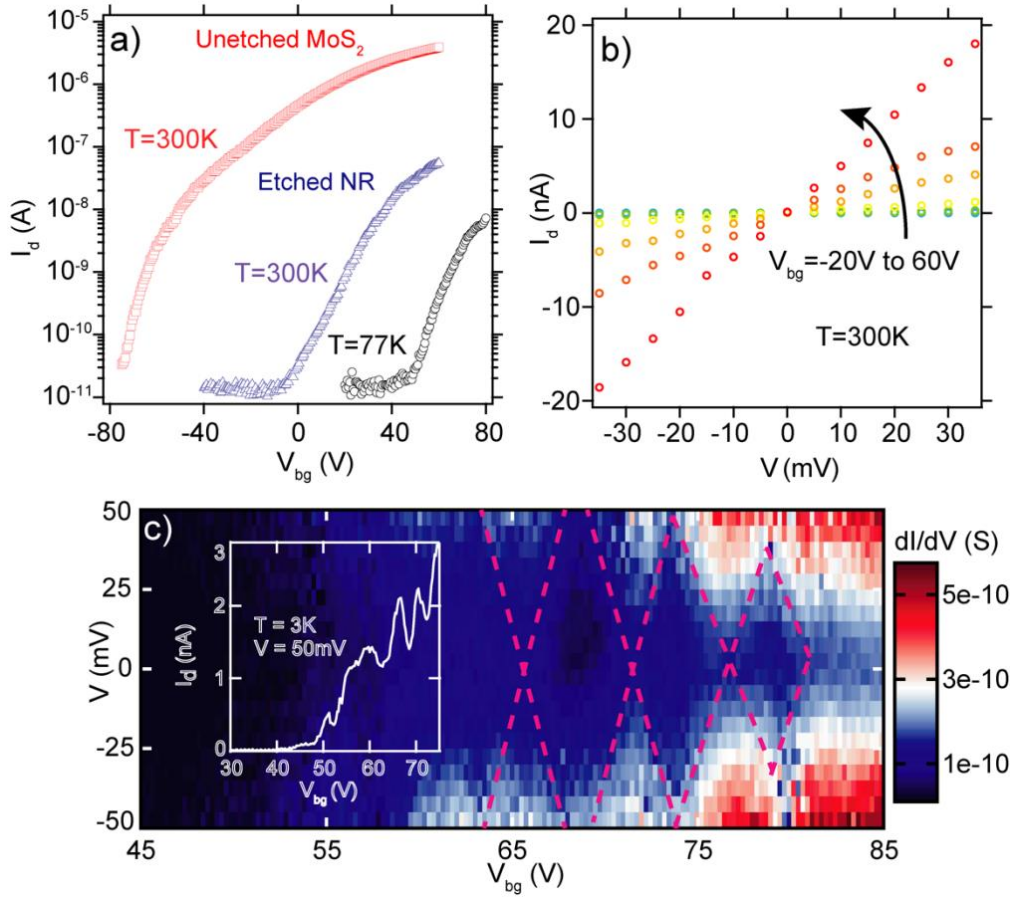

*Figure 2.a) Drain current ($I_d$) vs $V_{bg}$ at V=100mV for unetched MoS$_2$ flake at room temperature (•), etched NR at room temperature (Δ) and etched NR at T=77K (O). b) $I_d$ –V for varying $V_{bg}$ from -20V to 60V at room temperature exhibiting linear Ohmic behavior. c) 2D plot of $I_d$ as a function of V and $V_{bg}$ at T=3K. Current exhibits Coulomb diamonds for device D1 due to single electron tunneling through quantum dot. Inset in figure c) shows $I_d$ as a function of $V_{bg}$ at fixed V=50mV and T=3K.*

In addition to the shift in threshold voltage, the measured current through the NR dropped compared to unetched flake. This behaviour is expected since in a field effect transistor, channel resistance strongly depends on the width of the channel. Assuming linear dependence of $I_d$ on the width (W) of the device channel [22], the expected current for W=50nm is 35nA, close to the measured value [50nA at $V_{bg}$=60V for V=100mV] corresponding to a total resistance of 2MΩ. The field effect mobility extracted from the purple trace for NR at T=300K in figure 2a is 22cm$^2$/V.s. This increase in mobility in NR device compared to unetched flake is observed in other reports as well [23], but the phenomenon is not so well understood and needs further investigation which is beyond the scope of this current report. The transfer output characteristics in figure 2b shows linear $I_d$-V behaviour at different $V_{bg}$ = -20 to 60V. The increased room temperature mobility and linear transfer output measurements does not suggest a degradation of material quality at room temperature after etching the MoS$_2$ flake into a NR device. At T=77K, the NR device pinch-off voltage shifts further to the positive backgate voltage ($V_{bg}$ =40V) compared to the room temperature NR device pinch-off voltage and a decrease in current through the device is observed due to the suppression in thermally activated transport as is seen in other material systems as well [24].

At low temperature (T=3K), $I_d$ vs $V_{bg}$ (at V=50mV) exhibits oscillatory behaviour as shown in figure 2c inset. We attribute this to Coulomb oscillations due to single electron tunnelling through a quantum dot (origin of quantum dot discussed below). For $V_{bg}$ < 65V, no clear periodic oscillations are visible. However at $V_{bg}$ > 65V clear and regular $I_d$ oscillations emerge. Such $I_d$ oscillations may arise due to different phenomenon such as conductance quantisation [25,26], Fabry-Perot [27,28] or Coulomb blockade [12]. We rule out the possibility of conductance quantisation and Fabry-Perot interference mechanism because these mechanisms require the device to operate in (quasi-)ballistic regime with the carrier mean free path comparable to the channel length. In MoS$_2$, signatures of ballistic transport have been shown over a length approximately 200nm in a hBN encapsulated device [29]. The carrier mean free path in MoS$_2$ devices fabricated on SiO$_2$ substrate is expected to be much smaller than 200nm due to charge localisation in MoS$_2$ introduced by the defect states in SiO$_2$ substrate [31] (discussed below). Hence, we conclude that conductance quantisation and Fabry-Perot interference are unlikely to be the origin of such oscillations in our device. On the other hand, graphene on SiO$_2$ substrate or etched graphene NRs are known to introduce charge localisation resulting in Coulomb blockade. Extending similar arguments to our MoS$_2$ NR device, we argue for Coulomb blockade to be the origin of I$_d$ oscillations seen in figure 2c (see discussion section).

Figure 2c show the 2D I$_d$ map as a function of V and V$_{bg}$ at 3K forming Coulomb diamond for V$_{bg}$ > 65V. Complete blockade inside the Coulomb diamond is not seen in D1 due to finite transparency in barriers resulting in finite current in Coulomb blockade regime [12]. As seen from figure 2c, the size of the Coulomb diamond is decreasing with increase in V$_{bg}$. Since the Coulomb diamond size is related to the size of quantum dot, this suggests that the size of the quantum dot increases with increase in V$_{bg}$. From the slopes of the Coulomb diamond (e.g. first Coulomb diamond), we extract source and drain capacitances as 1.26aF and 1.23aF respectively and from the period of Coulomb oscillations we get the gate capacitance as 0.022aF with the total capacitance (C) of 2.52aF. The lever arm parameter ($\alpha$) which is the ratio of gate capacitance with total capacitance obtained is ~8.7meV/V which is comparable to the value reported with thick SiO$_2$ dielectrics [12]. Using the total capacitance, we calculate the charging energy, E$_c$(=$e^2$/C)=63meV which is consistent with the measured charging energy (first Coulomb diamond in figure 2c) verifying that the extracted capacitances are in the correct order of magnitude. Using C = 8$\varepsilon_0\varepsilon_r$R (where $\varepsilon_0$ is the permittivity of free space and $\varepsilon_r$ is the relative permittivity of SiO$_2$ ($\varepsilon_{r(SiO2)}$=3.9)) we estimate the quantum dot radius to be ~9nm.

Figure 3 shows low temperature measurements for D2 (few layer MoS$_2$ NR) with device dimensions of W=70nm and L=500nm at T=3K. 2D-map of I$_d$ (V$_{bg}$, V) in D2 exhibits clear Coulomb diamond as shown in figure 3a. Unlike in device D1, I$_d$ inside the Coulomb diamond in device D2 is significantly suppressed showing a strong Coulomb blockade effect. A line cut at the first Coulomb diamond crossing (green line in figure 3c) shows the $I_d$ peak when the quantum dot energy level is in resonance with source and drain electrochemical potential whereas $I_d$ is completely suppressed when the quantum dot energy level is shifted off resonance (inside the Coulomb diamond). Moreover, line cuts of $I_d$-V at two crossings of Coulomb diamond ($V_{bg}$= 13.3V and 16.3V) show linear increase in $I_d$ (figure 3c and 3e) whereas inside the Coulomb diamond, $I_d$-V shows suppression as expected ($V_{bg}$ = 14.7V, figure 3d). For the Coulomb blockade in classical regime, the $I_d$ lineshape is proportional to $\propto \text{Cosh}^{-2}(\alpha V_{bg}/3.5k_BT_e)$ [31], where $\alpha$ is the lever-arm parameter, $k_B$ is the Boltzmann constant and $T_e$ is the electron temperature. Fitting the $I_d$ line shape in figure 3b, we extract an electron temperature of 8.1K. Source, drain and gate capacitances extracted from Coulomb diamond in figure 3a are 3.27aF, 6.27aF and 0.06aF respectively with a total capacitance C=10aF and a lever arm parameter of 6meV/V. We find $E_c=e^2/C$=15.9meV, which is in good agreement with the measured value in figure 3a (~15meV). The radius of the quantum dot estimated from the total capacitance is ~36nm.

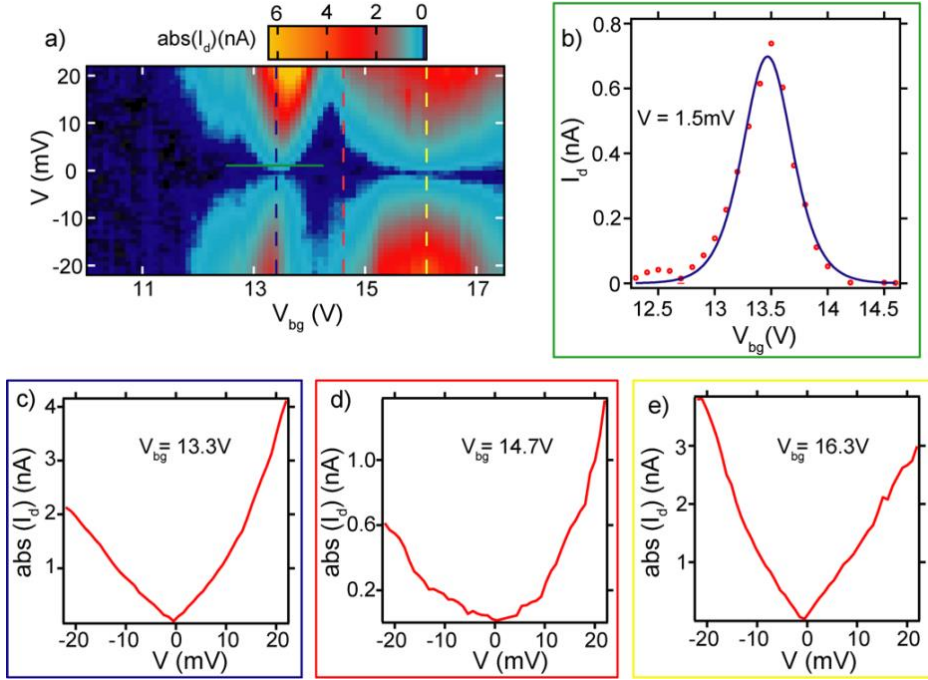

*Figure 3. Low temperature measurements of $I_d$ ($V_{bg}$, V) in device D2. a) $I_d$ ($V_{bg}$, V) at T=3K showing clear Coulomb diamond. b) $I_d$ linecut along $V_{bg}$ at a fixed V =1.5mV from data in panel (a) (green line). Blue curve is fitting to the Coulomb oscillation using $I_d \propto Cosh^{-2}(\alpha V_{bg}/3.5k_BT_e)$. c) and e) $I_d$-V line cuts at $V_{bg}$=13.3V (black line in panel a) and 16.3V (yellow line in panel a) when quantum dot energy level is in resonance with the source and drain electrochemical potential resulting in linear increase in $I_d$. d) Line cut of $I_d$-V at $V_{bg}$= 14.7V (red line in panel a) showing suppression of Id when quantum dot energy level is off-resonance with source and drain electrochemical potential.*

**DISCUSSION**

In this section, we discuss the two possible origins of quantum dot in our NR devices and ways to minimize or engineer the quantum dot. Firstly, the quantum dot can originate from the environment of the device which includes trap states (defects) in $SiO_2$ substrate [32], residues from fabrication process [33] and intrinsic defects in the $MoS_2$ lattice [34]. This results in charge localization in the $MoS_2$ NR (figure 4a) reflected as Coulomb oscillations at low temperature. Some of these defects can be largely reduced or eliminated by encapsulating the $MoS_2$ in hBN to minimize the influence of substrate as well as eliminate exposure to chemicals/ solvents from the fabrication process.

Secondly, the quantum dot may arise from the edge effects which includes microscopic roughness along the etched edges [35], molecule bound to edge [36] or edge reconfiguration [37]. Our NR device design was optimized to minimize long etched edges by defining the channel between the two metallic contacts separated by L=500nm compared to other device geometry where long etched edges were formed outside the channel area [38]. In our device, etching was performed only in a small region over the channel length is to minimize long edges. One solution to minimize this effect is to chemically functionalize the channel to generate smooth edges [36]. Alternatively, if the NR edges can be passivated by ALD oxide deposition on NR, this would facilitate utilizing local gates to smooth out the charge localization and tune out the unwanted states (shown in fig. 4c) and further confine the carriers in third dimension. In such a device with local gates, the size of the quantum dot is roughly defined by the cross-sectional overlap between the NR and metallic gate allowing a well-controlled quantum dot as shown in other material systems [39-41].

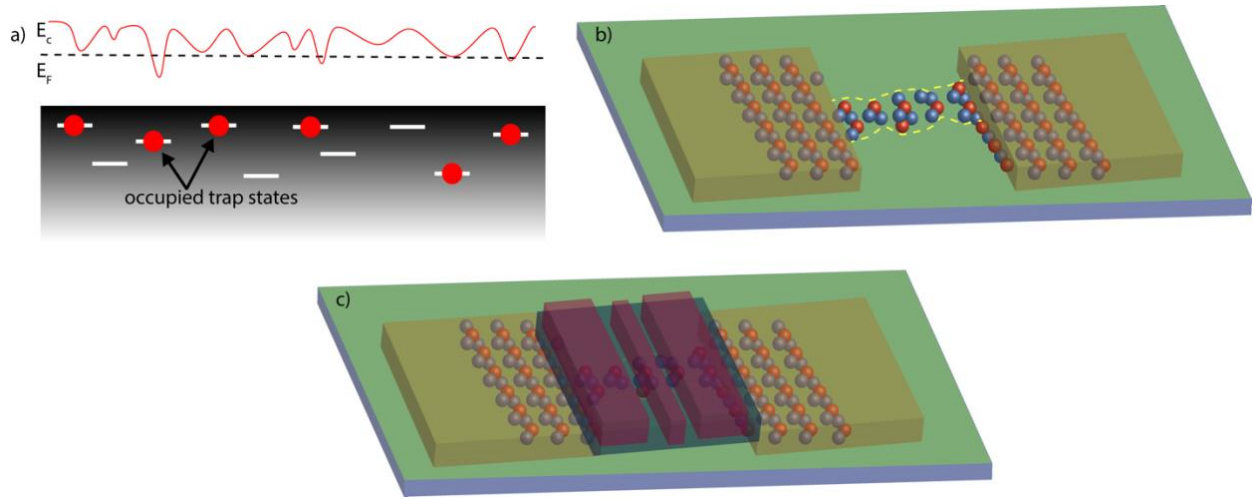

*Figure 4. a) Non-uniformity created in MoS$_2$ electrochemical potential due to trap states in SiO$_2$ substrate resulting in confinement. b) Artistic picture of an etched NR device. Non-uniformity due to rough edge could result in formation of a quantum dot outlined by yellow dotted line. c) Schematic of the NR device with local gates which can be possibly used to tune out the unwanted states to provide better control over defining the quantum dot in NR (light blue block on top of the NR region represents the dielectric layer and marron blocks are the local metallic gates).*

**CONCLUSION**

To conclude, we have performed the first low temperature transport spectroscopy of MoS$_2$ NR devices. We have successfully used etching to reduce one lateral dimension in monolayer and few layer MoS$_2$ and fabricated NRs that hosts quantum dots, as evidenced by Coulomb blockade at 3K. From our measurements, we estimate that these quantum dots have sizes below 50nm which is otherwise challenging using the split-gate technique. We demonstrate thus the possibility to employ nano-lithographic etching to achieve sub-50nm quantum dots. Future work will focus on further refinements of this technique using e.g. defect control, edge passivation by chemical functionalization or ALD oxide, etc. Additionally, by employing local gating one could produce quantum dots which are completely defined by lithography. This paves the way for spin based quantum information processing in 2D materials.

AUTHOR INFORMATION

**Corresponding Author**

*patild@imre.a-star.edu.sg

*dharamkotekar@gmail.com

§kejgoh@yahoo.com

**Author Contributions**

DKP designed the experiments. DKP fabricated the NR devices with assistance from JD. SLW performed optical characterization of the MoS$_2$ flakes. DKP performed the measurements and analyzed the data. DKP and KEJG lead the project. All the authors discussed and wrote the manuscript.

**Funding Sources**

This work was supported by the A*STAR (Singapore) QTE Grant No. A1685b0005.

ACKNOWLEDGMENT

We acknowledge Ooi Zi En for assistance in 3K-cryogenic set-up.

REFERENCES

1. Rajesh Kappera, Damien Voiry, Sibel Ebru Yalcin, Brittany Branch, Gautam Gupta, Aditya D. Mohite & Manish Chhowalla, Phase-engineered low-resistance contacts for ultrathin $MoS_2$ transistors, Nature Materials volume13, pages1128–1134 (2014).

2. Das, S., Chen, H.Y., Penumatcha, A. V. & Appenzeller, J. High performance multilayer MoS2 transistors with scandium contacts. Nano Lett. 13, 100–105 (2013).

3. Dankert, A., Langouche, L., Kamalakar, M. V. & Dash, S. P. High-performance molybdenum disulfide field-effect transistors with spin tunnel contacts. ACS Nano 8, 476–482 (2014).

4. Maurel, C., Coratger, R., Ajustron, F., Beauvillain, J. & Gerard, P. Electrical characteristics of metal/semiconductor nanocontacts using light emission in a scanning tunneling microscope. J. Appl. Phys. 94, 1979–1982 (2003).

5. Gourmelon, E., Bernède, J. C., Pouzet, J. & Marsillac, S. Textured MoS2 thin films obtained on tungsten: Electrical properties of the W/MoS2 contact. J. Appl. Phys. 87, 1182–1186 (2000).

6. Leong, W. S. et al. Low resistance metal contacts to MoS2 devices with nicketched-graphene electrodes. ACS Nano 9, 869–877 (2015).

7. Adrien Allain, Jiahao Kang, Kaustav Banerjee & Andras Kis, Electrical contacts to two-dimensional semiconductors, Nature Materials volume 14, pages 1195-1205 (2015).

8. Riccardo Pisoni, Yongjin Lee, Hiske Overweg, Marius Eich, Pauline Simonet, Kenji Watanabe, Takashi Taniguchi, Roman Gorbachev, Thomas Ihn, and Klaus Ensslin, Gate-Defined One-Dimensional Channel and Broken Symmetry States in $MoS_2$ van der Waals Heterostructures, Nano Lett. 2017, 17, 8, 5008-5011 (2017).

9. Alexander Epping, Luca Banszerus, Johannes Güttinger, Luisa Krückeberg, Kenji Watanabe, Takashi Taniguchi, Fabian Hassler, Bernd Beschoten and Christoph Stampfer, Quantum transport through MoS2 constrictions defined by photodoping, Journal of Physics: Condensed Matter, Volume 30, Number 20 (2018).

10. Riccardo Pisoni, Zijin Lei, Patrick Back, Marius Eich, Hiske Overweg, Yongjin Lee, Kenji Watanabe, Takashi Taniguchi, Thomas Ihn, and Klaus Ensslin, Gate-Tunable Quantum Dot in a High Quality Single Layer MoS2 Van der Waals Heterostructure, Applied Physics Letters 112, 123101 (2018).

11. Ke Wang, Kristiaan De Greve, Luis A. Jauregui, Andrey Sushko, Alexander High, You Zhou, Giovanni Scuri, Takashi Taniguchi, Kenji Watanabe, Mikhail D. Lukin, Hongkun Park and Philip Kim, Electrical control of charged carriers and excitons in atomically thin materials, Nature Nanotechnology volume 13, pages128–132 (2018).

12. Kyunghoon Lee, Girish Kulkarni and Zhaohui Zhong, Coulomb blockade in monolayer MoS2 single electron transistor, Nanoscale, 2016, 8, 7755-7760 (2016).

13. Xiang-Xiang Song, Di Liu, Vahid Mosallanejad, Jie You, Tian-Yi Han, Dian-Teng Chen, Hai-Ou Li, Gang Cao, Ming Xiao, Guang-Can Guo and Guo-Ping Guo, A gate defined quantum dot on the two-dimensional transition metal dichalcogenide semiconductor WSe2, *Nanoscale*, 2015, 7, 16867 – 16873 (2015).

14. Zhuo-Zhi Zhang, Xiang-Xiang Song, Gang Luo, Guang-Wei Deng, Vahid Mosallanejad, Takashi Taniguchi, Kenji Watanabe, Hai-Ou Li, Gang Cao, Guang-Can Guo, Franco Nori and Guo-Ping Guo, Electrotunable artificial molecules based on van der Waals heterostructures, Science Advances 20 Oct 2017: Vol. 3, no. 10 (2017).

15. Xiang-Xiang Song, Zhuo-Zhi Zhang, Jie You, Di Liu, Hai-Ou Li, Gang Cao, Ming Xiao and Guo-Ping Guo, Temperature dependence of Coulomb oscillations in a few-layer two-dimensional $WS_2$ quantum dot, Scientific Reports volume 5, Article number: 16113 (2015).

16. Khairul Alam and Roger K. Lake, Monolayer MoS2 Transistors Beyond the Technology Road Map, IEEE Transactions on Electron Devices Volume 59 , Issue 12 , Dec. (2012).

17. Han Liu, Mengwei Si, Yexin Deng, Adam T. Neal, Yuchen Du, Sina Najmaei, Pulickel M. Ajayan, Jun Lou, and Peide D. Ye, Switching Mechanism in Single-Layer Molybdenum Disulfide Transistors: An Insight into Current Flow across Schottky Barriers, ACS Nano, 2014, 8 (1), pp 1031–1038 (2013).

18. A. Splendiani, L. Sun, Y. Zhang, T. Li, J. Kim, C.-Y. Chim, G. Galli, and F. Wang, Emerging Photoluminescence in Monolayer MoS2, Nano Letters 10, 1271 (2010).

19. Y.-H. Lee, X.-Q. Zhang, W. Zhang, M.-T. Chang, C.- T. Lin, K.-D. Chang, Y.-C. Yu, J. Tse-Wei Wang, C.-S. Chang, L.-J. Li, and T.-W. Lin, Synthesis of Large-Area

MoS$_2$ Atomic Layers with Chemical Vapor Deposition, Advanced Materials Volume 24, Issue17, 2320-2325 (2012).

20. S. Wu, C. Huang, G. Aivazian, J. S. Ross, D. H. Cobden, and X. Xu, ACS Nano 7, 2768 (2013).

21. W. Liu, D. Sarkar, J. Kang, W. Cao, and K. Banerjee, Impact of Contact on the Operation and Performance of Back-Gated Monolayer MoS2 Field-Effect-Transistors, ACS Nano 9, 7904 (2015).

22. H. Liu, J. Gu, and P. D. Ye, MoS2 Nanoribbon Transistors: Transition From Depletion Mode to Enhancement Mode by Channel-Width Trimming, IEEE Electron Device Letters 33, 1273 (2012).

23. D. Kotekar-Patil, J. Deng, S. L. Wong, Chit Siong Lau, and Kuan Eng Johnson Goh, Single layer MoS2 nanoribbon field effect transistor, Applied Physics Letters 114, 013508 (2019).

24. Subhash C. Rustagi, N. Singh, Y. F. Lim, G. Zhang, S. Wang, G. Q. Lo, N. Balasubramanian, and D.-L. Kwong, Low-Temperature Transport Characteristics and Quantum-Confinement Effects in Gate-All-Around Si-Nanowire N-MOSFET, IEEE ELECTRON DEVICE LETTERS, VOL. 28, NO. 10, OCTOBER (2007).

25. Endre Tóvári, Péter Makk, Ming-Hao Liu, Peter Rickhaus, Zoltán Kovács-Krausz, Klaus Richter, Christian Schönenberger and Szabolcs Csonka, Gate-controlled conductance enhancement from quantum Hall channels along graphene p–n junctions, Nanoscale, 2016, 8, 19910-19916 (2016).

26. M. J. Biercuk, N. Mason, J. Martin, A. Yacoby, and C. M. Marcus, Anomalous Conductance Quantization in Carbon Nanotubes, Phys. Rev. Lett. 94, 026801 (2005).

27. Wenjie Liang, Marc Bockrath, Dolores Bozovic, Jason H. Hafner, M. Tinkham and Hongkun Park, Fabry - Perot interference in a nanotube electron waveguide, Nature volume411, pages665–669 (07 June 2001).

28. D. Kotekar-Patil, B.-M. Nguyen, J. Yoo, S. A. Dayeh, and S. M. Frolov, Quasiballistic quantum transport through Ge/Si core/shell nanowires, Nanotechnology 28, 385204 (2017).

29. Riccardo Pisoni, Yongjin Lee, Hiske Overweg, Marius Eich, Pauline Simonet, Thomas Ihn, and Klaus Ensslin, Gate-Defined One-Dimensional Channel and Broken Symmetry States in MoS2 van der Waals Heterostructures, *Nano Letters*, 2017, 17 (8), pp 5008–5011 (2017).

30. Subhamoy Ghatak, Atindra Nath Pal, and Arindam Ghosh, The Nature of Electronic States in Atomically Thin MoS2 Field-Effect Transistors, *ACS Nano*, 2011, 5 (10), pp 7707–7712 (2011).

31. C. W. J. Beenakker, Theory of Coulomb-blockade oscillations in the conductance of a quantum dot, Physical Review B 44, 1646 (1991).

32. J. Xue, J. Sanchez-Yamagishi, D. Bulmash, P. Jacquod, A. Deshpande, K. Watanabe, T. Taniguchi, P. Jarillo-Herrero, and B. J. LeRoy, Scanning tunnelling microscopy and spectroscopy of ultra-flat graphene on hexagonal boron nitride, Nature Mater. 10, 282–285 (2011).

33. J. Moser, A. Barreiro, and A. Bachtold, Current-induced cleaning of graphene, Appl. Phys. Lett. 91, 163513 (2007).

34. J.-H. Chen, W. G. Cullen, C. Jang, M. S. Fuhrer, and E. D. Williams, Defect Scattering in Graphene, Phys. Rev. Lett. 102, 236805 (2009).

35. D. Bischoff, A. Varlet, P. Simonet, M. Eich, H. C. Overweg, T. Ihn, and K. Ensslin, Localized charge carriers in graphene nanodevices, Applied Physics Reviews 2, 031301 (2015).

36. T. Kato, L. Jiao, X. Wang, H. Wang, X. Li, L. Zhang, R. Hatakeyama, and H. Dai, Room-Temperature Edge Functionalization and Doping of Graphene by Mild Plasma, Small 7, 574–577 (2011).

37. X. Jia, M. Hofmann, V. Meunier, B. G. Sumpter, J. Campos-Delgado, J. M. Romo-Herrera, H. Son, Y.-P. Hsieh, A. Reina, J. Kong, M. Terrones, and M. S. Dresselhaus, Controlled formation of sharp zigzag and armchair edges in graphitic nanoribbons, Science 323, 1701–1705 (2009).

38. D. Bischoff, F. Libisch, J. Burgdörfer, T. Ihn, and K. Ensslin, Characterizing wave functions in graphene nanodevices: Electronic transport through ultrashort graphene constrictions on a boron nitride substrate, Phys. Rev. B 90, 115405 (2014).

39. S. Nadj-Perge, S. M. Frolov, E. P. A. M. Bakkers and L. P. Kouwenhoven, Spin–orbit qubit in a semiconductor nanowire, *Nature* volume468, pages1084–1087 (2010).

40. X. Jehl, B. Roche, M. Sanquer, B. Voisin, R. Wacquez, V. Deshpande, et. al., Mass Production of Silicon MOS-SETs: Can We Live with Nano-Devices' Variability? Procedia Computer Science 7, (2011), Pages 266–268.

41. D Kotekar-Patil, A Corna, R Maurand, A Crippa, A Orlov, S Barraud, X Jehl, S De Franceschi, M Sanquer, Pauli spin blockade in CMOS double quantum dot devices, Phys. Status Solidi B 254, No. 3, 1600581 (2017).